\documentclass[aps,pra,twocolumn,showpacs,floatfix,superscriptaddress]{revtex4}
\usepackage{times,graphicx,amsfonts,amsmath,amssymb,color}
\newcommand{\quotes}[1]{``#1''}
\def\sb{{\scriptscriptstyle B}}
\def\ssg{{\scriptscriptstyle G}}
\def\sg{{\scriptscriptstyle N\!G}}
\def\sv{{\scriptscriptstyle V}}
\def\sh{{\scriptscriptstyle H}}
\def\sr{{\scriptscriptstyle R}}
\def\sm{{\scriptscriptstyle M}}
\def\smpt{{\scriptscriptstyle P}}
\def\stpt{{\scriptscriptstyle T}}
\def\sfs{{\scriptscriptstyle F}}
\newcommand{\braket}[2]{\langle #1|#2\rangle}

\newcommand{\Tr}{\hbox{Tr}}
\begin{document}
\title{Quantifying the nonlinearity of a quantum oscillator}
\author{Matteo G. A. Paris}
\affiliation{Dipartimento di Fisica, Universit\`a degli Studi di Milano, I-20133
Milano, Italy}
\author{Marco G. Genoni}
\affiliation{Department of Physics \& Astronomy, University College London, 
London WC1E 6BT, United Kingdom}
\author{Nathan Shammah}
\affiliation{
School of Physics and Astronomy, University of Southampton, Southampton,
SO17 1BJ, United Kingdom}
\author{Berihu Teklu}
\affiliation{Dipartimento di Fisica, Universit\`a degli Studi di Milano, I-20133
Milano, Italy}
\begin{abstract}
We address the quantification of nonlinearity for quantum oscillators
and introduce two measures based on the properties of the ground state
rather than on the form of the potential itself.  The first measure is a
fidelity-based one, and corresponds to the renormalized Bures distance
between the ground state of the considered oscillator  and the ground
state of a reference harmonic oscillator.  Then, in order to avoid the
introduction of this auxiliary oscillator, we introduce a different
measure based on the non-Gaussianity (nG) of the ground state. The two
measures are evaluated for a sample of significant nonlinear potentials
and their properties are discussed in some detail.  We show that the two
measures are monotone functions of each other in most cases, and this
suggests that the nG-based measure is a suitable choice to capture the
anharmonic nature of a quantum oscillator, and to quantify its
nonlinearity independently on the specific features of the potential.
We also provide examples of potentials where the Bures measure cannot be
defined, due to the lack of a proper reference harmonic potential, while
the nG-based measure properly quantify their nonlinear features. Our
results may have implications in experimental applications where access
to the effective potential is limited, e.g., in quantum control, and
protocols rely on information about the ground or thermal state.
\end{abstract}
\date{\today}
%%%%
\maketitle
%%%%
\section{Introduction}
%%%%
Oscillators represent one of the main conceptual and technical tools in
physics. At a quantum level, oscillators capture, e.g., the physics of
light trapped in a cavity, the nature of molecule bonding, and the
behavior of optomechanical oscillators, cantilevers or springs.  As a
matter of fact, any bounded quantum system may be always described as a
quantum harmonic oscillator (QHO) after a suitable approximation. At the
same time, nonlinear features are relevant in many fields of physics and
they can be exploited for several applications \cite{nlrv,luk94}. As 
for example, nonlinearity has been recently exploited in optomechanical 
systems to generate single-photon states and more general non-Gaussian states
\cite{nnk11}. Nonlinear oscillators attracted attention 
also from a purely mathematical point of view, and may have potential
applications in different areas also outside physical sciences.
\par 
Quantum technology in continuous-variable systems has been initially
developed with Gaussian states \cite{revg1,revg2,revg3,revg4}. More
recently, however, the use of non-Gaussian states and operations
\cite{ng1,ng2,ng3,ng4} has emerged as a resource for enhancing several 
processes, as entanglement distillation \cite{nGDist1,nGDist2,nGDist3}, 
quantum estimation \cite{geoNG}, 
and quantum error correction \cite{nGEC}.
In this framework, nonlinear oscillators may be useful since their
ground states (GSs), as well as states at thermal equilibrium, are
necessarily non-Gaussian.  In the realm of discrete variables, nonlinear
oscillators allow to engineer effective two-level systems, due to their
varying spacing of its energy levels in contrast to uniform (harmonic)
spacing.  In fact, strategies to generate entanglement with these
effective qubits have been already proposed \cite{owen}.  
\par 
The above arguments suggest that nonlinearity may represent a
resource for quantum information and control. In order to investigate
whether this is indeed the case, one needs to quantify the degree of
nonlinearity of a given system, i.e. to introduce a measure 
for the nonlinear character of a quantum oscillator. On a more 
fundamental perspective, it would be of interest to have a measure 
of quantum nonlinearity in order to investigate quantitatively the 
connection between the rate of decoherence of an open quantum system 
and the anharmonic features of the potential \cite{Vie11}.
\par
"What is the nonlinearity of an oscillator?" is a 
seemingly simple question, which however is not immediate to
address in the quantum case and, in turn, did not receive a general 
answer so far.
Our approach to the problem is to focus on the properties of the 
oscillators' ground state rather than on the specific features 
of the potential itself. Indeed, it would be desirable to define 
a measure that assesses the physical effects of the nonlinearity, rather than the 
dependence on the parameters that appear in the expression of the 
potentials. In addition, the potential functions are, in general, 
not integrable functions on the real axis and this, loosely speaking,  
prevents any attempt to introduce a nonlinearity measure based 
on any distance function between potentials. We thus shift our attention 
from the potential to the ground state associated to that potential and 
compare their properties to the ground state of the harmonic oscillator.
\par
In particular, we focus on the Gaussian character of the harmonic 
oscillator GS and introduce a nonlinearity measure based on the
non-Gaussianity of the GS. We analyze in some detail the properties 
of this measure and compare its behavior with that of another possible 
choice, i.e. the Bures distance between the potential's GS
and its QHO counterpart. Although the Bures-based 
nonlinearity measure $\eta_\sb \left[\hbox{V}\right]$ somehow 
represents a \quotes{natural} choice, it requires the introduction
of a reference harmonic oscillator, i.e. the knowledge of the potential
function at least in the vicinity of its minimum. On the contrary, 
the non-Gaussianity-based (nG-based) measure 
$\eta_\sg\left[\hbox{V}\right]$ only requires the knowledge of the GS, 
and thus it may represent a more convenient choice for experimental
applications. As we will see, the two measures are monotone functions of 
each other in most cases, and this suggests that the nG-based measure is a 
convenient choice to capture the anharmonic nature of quantum
oscillators and to quantify their nonlinearity in a robust way, e. g., 
independently on the specific features of the potential. 
In addition, we will also provide examples of potentials where the Bures measure
cannot be defined, due to the lack of a proper reference harmonic
potential, while the nG-based measure properly quantifies their 
nonlinear features. 
\par
The rest of the paper is structured as follows: In the next section we 
review few facts about quantum harmonic oscillators, in order to establish
notation, and introduce the two nonlinearity measures, also discussing
their general properties. In section \ref{s:pots} we present
results for a choice of significant nonlinear potentials and 
discuss their implications. Section \ref{s:out} closes the paper with 
some concluding remarks.
%%%%
\section{Nonlinearity of a quantum oscillator}
%2
The potential of the quantum harmonic oscillator, 
$V_\sh (x)=\frac{1}{2}m\omega^2x^2$ is fully characterized by 
its frequency, provided that the 
mass of the oscillator is normalized to unity, $m=1$.
The Hamiltonian of the QHO describes its energy, and is given by
$H=\frac12 p^2 +\frac{1}{2}\omega^2x^2$ 
where $p$ and $x$ are the momentum and position operators. 
One can introduce the ladder operators, $a$ and $a^\dag$,
for which $\left[ a, a^\dagger \right] =\mathbb{I}$,
so that the Hamiltonian can be rewritten as
$H=\omega \left(a^\dag a +\frac{1}{2}\right)$, 
with $a^\dag a=\hat{N}$ representing the number operator. The energy spectrum is given by 
$E_n=\omega\left(n+\frac{1}{2}\right)$ 
with $n=\{0,1,2,...\}$. It is lower bounded, discrete, infinite, and equally spaced. 
By projecting the eigenstates onto the position 
basis, one obtains the $n$-th eigenfunction
in terms of the Hermite polynomial, ${\mathcal{H}}_n\left(z\right)$. 
As $\mathcal{H}_{0}(x)=1$,
the GS of the QHO is described by the Gaussian wavefunction,
\begin{equation}
\psi_\sh(x)=\braket{x}{0}_\sh =\left(\frac{\omega}{ 
\pi}\right)^{\frac{1}{4}}e^{-\frac12\omega x^2}\,.
\end{equation}
In order to quantify the nonlinearity of a quantum oscillator we
compare the GS of the considered potential with the GS of the 
harmonic oscillator. \par
The first measure that we put forward 
involves the comparison in terms of fidelity.
The general expression for the fidelity between two quantum states, 
$\rho$ and $\tau$, is given by
$F[\rho,\tau]=\hbox{Tr}\left[\sqrt{\sqrt{\tau}\rho\sqrt{\tau}}\right]^2$. 
For pure states $\rho=|\psi\rangle\langle\psi |$ and 
$\tau=|\phi\rangle\langle\phi |$, the formula reduces to the 
overlap $F[\rho,\tau]=|\langle\phi | \psi\rangle|^2$. 
The Bures distance is given by 
\begin{equation}
D_{\text{B}}[\rho,\tau]=\sqrt{2\left(1-\sqrt{F[\rho,\tau]} \right)}\,.
\label{eq:fidme}
\end{equation}
Given a potential $V(x)$ leading to an oscillatory behavior, we define the 
nonlinearity measure $\eta_\sb\left[\hbox{V}\right]$ as the 
renormalized Bures distance between the GS $|0\rangle_\sv$ of the
quantum oscillator under consideration and the GS $| 0\rangle_\sh$ 
of the corresponding harmonic oscillator 
\begin{equation}
\eta_\sb\left[\hbox{V}\right]=\frac1{\sqrt{2}}\,D_\sb \left[| 0\rangle_\sv,
|0\rangle_\sh
\right]\,.
\label{eq:etaA}
\end{equation} 
Since the two GSs are pure states then 
$\eta_\sb\left[\hbox{V}\right]$ may be written 
as \begin{equation}
 \eta_\sb\left[\hbox{V}\right]=\sqrt{1-|{}_\sh\langle 0
 |0\rangle_\sv| }\,.
 \label{etaB}
 \end{equation}
As it is apparent from its very definition, this measure of nonlinearity
depends upon the choice of a corresponding {\em reference harmonic
oscillator}.  By this we mean the harmonic oscillator with a frequency
$\omega_\sr$ that approximates the nonlinear potential, for small
displacement, i.e.  in the vicinity of its minimum $V(x)\simeq
\frac{1}{2}\omega_\sr^{2}x^{2}$ (assuming a reference system centered
at the potential minimum). Here $\omega_\sr$ is a function of the
nonlinear parameters appearing in the functional form of the potential
$V(x)$, i.e. $\omega_\sr=\omega_\sr \left(\alpha_{1},\alpha_{2},...\right)$. 
This is a quite natural choice for the reference oscillator. However,
depending on the potential under investigation, the determination of 
this frequency may be problematic or even misleading.  
\par
This issue leads us to
introduce a different measure, $\eta_\sg [V]$, which does not depend 
on the choice of a reference potential.
In fact, given a potential $\hbox{V}(x)$, an alternative definition for
a nonlinearity measure $\eta_\sg \left[\hbox{V}\right]$, may be given in
terms of the non-Gaussianity of the corresponding GS, i.e. 
\begin{equation}
\eta_\sg \left[\hbox{V}\right]=\delta_\sg
\left[| 0\rangle_\sv {}_\sv \langle 0|  \right]
\label{etaNG}
\end{equation}
where $ \delta_\sg[\varrho ]$ is the non-Gaussianity measure 
introduced in 
\cite{ng2,ng3}, built on the quantum relative entropy of the state
and a reference Gaussian state, with the procedure hereby reviewed. 
Given two quantum states $\rho$ and $\tau$, the quantum relative
entropy (QRE) is defined as \begin{equation}
S\left(\rho ||\tau\right)=
\text{Tr}\left[\rho\right(\ln\rho-\ln \tau\left)\right]\,.
\end{equation}
Despite not being strictly a
distance in the mathematical sense (it is not symmetric and does not
obey a triangle inequality), the QRE is always non-negative and in
particular one has $S(\rho||\tau) = 0$ iff $\rho= \tau$.
Moreover, it has  a nice operational interpretation in terms of
distinguishability of quantum states: given two quantum states $\rho$ 
and $\tau$, the probability $P_N$ that the state $\tau$ is confused 
with $\rho$ after $N$ measurements is 
$P_{N} =\exp\{-N S\left(\rho || \tau,\right)\}$, 
as $N\rightarrow \infty$. This further supports the view of the QRE as a
distance-like quantity between quantum states in the Hilbert space.
Among its properties, we mention that the QRE is invariant under unitary
operations and not-increasing under generic quantum maps. The QRE-based 
measure of nG is defined as \cite{ng2,ng3}
\begin{equation}
\delta\sg[\rho]=S\left(\rho ||\tau_\ssg\right)
\label{eq:qrenG}
\end{equation}
where $\tau_\ssg$ is the reference Gaussian state of $\rho$, i.e. 
a Gaussian state with the same covariance matrix of the state 
$\rho$ (see Appendix \ref{apb}). For single-mode states we have
\begin{equation}
\delta_\sg[\rho]=S(\tau_\ssg)-S(\rho)=h(\sqrt{\hbox{det}[\sigma}])-S(\rho)
\label{eq:ng}
\end{equation}
where 
$h(x)=(x+\frac{1}{2})\ln(x+\frac{1}{2})-(x-\frac{1}{2})\ln(x-\frac{1}{2})$
and  $S(\rho)=-\hbox{Tr}\left[\rho\,\ln \rho\right]$ is the 
Von Neumann entropy of the state. 
For pure states, $S(\rho)=0$, and thus
\begin{equation}
\delta_\sg [|\psi\rangle\langle\psi |]=
h\left(\sqrt{\hbox{det}[\sigma]}\right)\,,
\label{eq:ng2}
\end{equation}
where $\sigma$ is the covariance matrix of the GS, built
using the first moments of the canonical operators.
The crucial
point here is that the definition of $\eta_\sg$ requires the determination
of a {\em reference Gaussian state} for the GS of $V(x)$ rather than a
{\em reference harmonic potential} for $V(x)$ itself. 
Therefore, whereas the calculation of $\eta_\sb$ requires 
the knowledge of the 
behavior of the potential near its minimum, the nonlinearity measure
$\eta_\sg$ is independent on the specific features of the potential.
This property is particularly relevant for possible experimental applications
of these measures: the GS wavefunction of a given nonlinear 
potential can be indeed tomographically estimated \cite{gentomo} and, 
as a consequence, the measure 
$\eta_\sg$ can be directly evaluated without any {\em a priori} information 
on the potential. 
{In addition 
the measure $\eta_\sg$ inherits an important property from the non-Gaussianity
measure $\delta_\sg$, which helps in justifying its use for quantifying nonlinearity
from a more mathematical and fundamental point of view.
As it is proved in \cite{ng2}, $\delta_\sg$ is invariant under symplectic transformations, 
i.e. transformation induced by any Hamiltonian which is  
quadratic or linear in the field operators (or in the canonical coordinates).  
It is clear then that its property induces an expected and more than 
reasonable hierarchy between Hamiltonians by means of the measure $\eta_\sg$, 
assigning the same amount of nonlinearity for all the Hamiltonians which 
are related by a  symplectic transformation, e.g. oscillators 
that are simply displaced one from each other, rotated in phase-space, or even squeezed.}
%%%%
\subsection{Properties of the nonlinearity measures}
Before presenting some examples illustrating the behavior of $\eta_\sg$ 
and $\eta_\sb$ for specific nonlinear potentials, let us describe some
general properties of the nonlinearity measures we have just introduced. 
Both measures are zero for a harmonic potential, whereas they may lead 
to a different definition of maximally nonlinear potential.
\par
The Bures-based nonlinearity is a bounded function, $0\leq \eta_\sb\leq 1$. The
maximum is achieved for potentials which have a GS
orthogonal to that of the corresponding harmonic oscillator, e.g., Fock
number states or any other state residing in the subspace orthogonal to
the \quotes{harmonic vacuum}. On the other hand, the nG-based nonlinearity 
$\eta_\sg\in [0,\infty)$ is an unbounded function. 
A renormalized quantity in $[0,1]$ may be obtained at any fixed value of 
energy upon normalizing $\eta_\sg$ to the non-Gaussianity of the 
maximally non-Gaussian state at that value of the energy, e.g., Fock 
number states. The maximum is thus achieved for a potential which has 
a GS equal to a Fock state of the harmonic oscillator or to
some specific superpositions of them \cite{ng3}.
\par
In order to gain some more insight into the properties of the two measures let
us consider a one-dimensional oscillatory system, whose harmonic behavior is  
perturbed by an anharmonic term in the potential, i.e. 
$$
V(x) = \frac12\, \omega^2 x^2 + \epsilon\, U (x)\,.
$$
According to standard perturbation theory for static Hamiltonians the 
ground state of the system may be approximated by
\begin{align}
\label{gsv}
|0\rangle_\sv &= 
|0\rangle_\sh - \epsilon  \sum_{k\neq 0} \frac{U_{k0}}{k} |k\rangle_\sh
\\ \notag 
&\simeq N^{-\frac12} \left ( |0\rangle_\sh + \alpha_1
|1\rangle_\sh + \alpha_2 |2\rangle_\sh \right)\,,
\end{align}
where $N=1+\alpha_1^2+\alpha_2^2$, 
$\alpha_k = -\epsilon U_{k0}/k$ and $U_{k0}={}_\sh\langle
k|U| 0 \rangle_\sh$, $k=1,2$. We retained the first two terms
in the perturbation expansion in order to describe situations where we
have some symmetries in the potential. Indeed, since the harmonic ground
state has an even wavefunction, an even anharmonic perturbation $U(x)$
would lead to $U_{10}=0$, whereas for an odd one we would have
$U_{20}=0$ (of course, a purely odd perturbation is not allowed, since
it would make the whole Hamiltonian unbounded from below). 
Upon expanding the perturbing potential to the fourth order, i.e. 
$$\epsilon U(x) \simeq \epsilon_3 x^3 + \epsilon_4
x^4\,,$$ we may write
\begin{align}
\label{alphas}
\alpha_1 &= - \epsilon_3\, {}_\sh\langle 1 | x^3 |0 \rangle_\sh = 
-\frac{3\epsilon_3}{(2 \omega)^\frac32} \\
\alpha_2 &= -\frac12 \epsilon_4\, {}_\sh\langle 2 | x^4 |0 \rangle_\sh
= - \frac12\epsilon_4 \frac{3}{\sqrt{2}}\frac1{\omega^2}\,. \notag
\end{align}
Within these assumptions, the nonlinearity measure $\eta_\sb$ can be evaluated 
in a straightforward way using Eqs. (\ref{etaB}), leading  to
$$
\eta_\sb [V]=\sqrt{1-N^{-\frac12}}\,.
$$
The nonlinearity measure $\eta_\sg[V]$ is obtained straightforwardly 
by  noticing that no correlations are present in the GS $|0\rangle_\sv$, such 
that $\det[\sigma] =
\overline{\Delta q^2}\,\overline{\Delta p^2}$
is given by the uncertainty product of the canonical operators, where
\begin{align}
\overline{\Delta q^2} & = \frac1{2N^2} \Big[3 \alpha_1^4 - 6 \sqrt{2}
\alpha_1^2 \alpha_2 \notag \\ & \qquad\qquad\;\;\quad 
+ (1+\alpha_2^2)(1+2\sqrt{2} \alpha_2 + 5 \alpha_2^2)\Big]
\,,\notag \\
\overline{\Delta p^2} & = \frac32 - \frac{1}{N} \left(1+\sqrt{2}
\alpha_2 - \alpha_2^2\right)\,.
\end{align}
%%%%%%
\begin{figure}[h!]
\includegraphics[width=0.42\columnwidth]{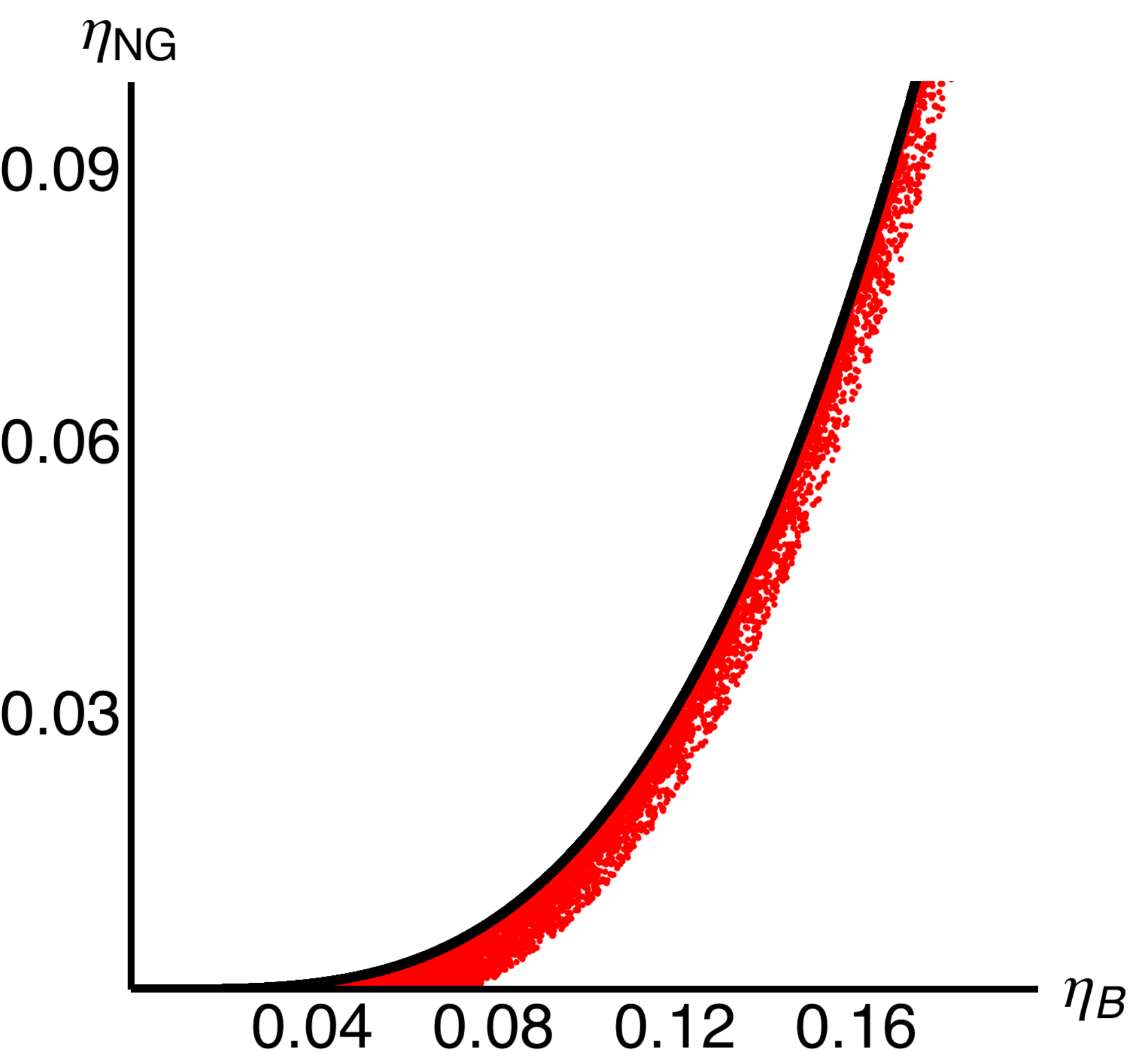}
\includegraphics[width=0.42\columnwidth]{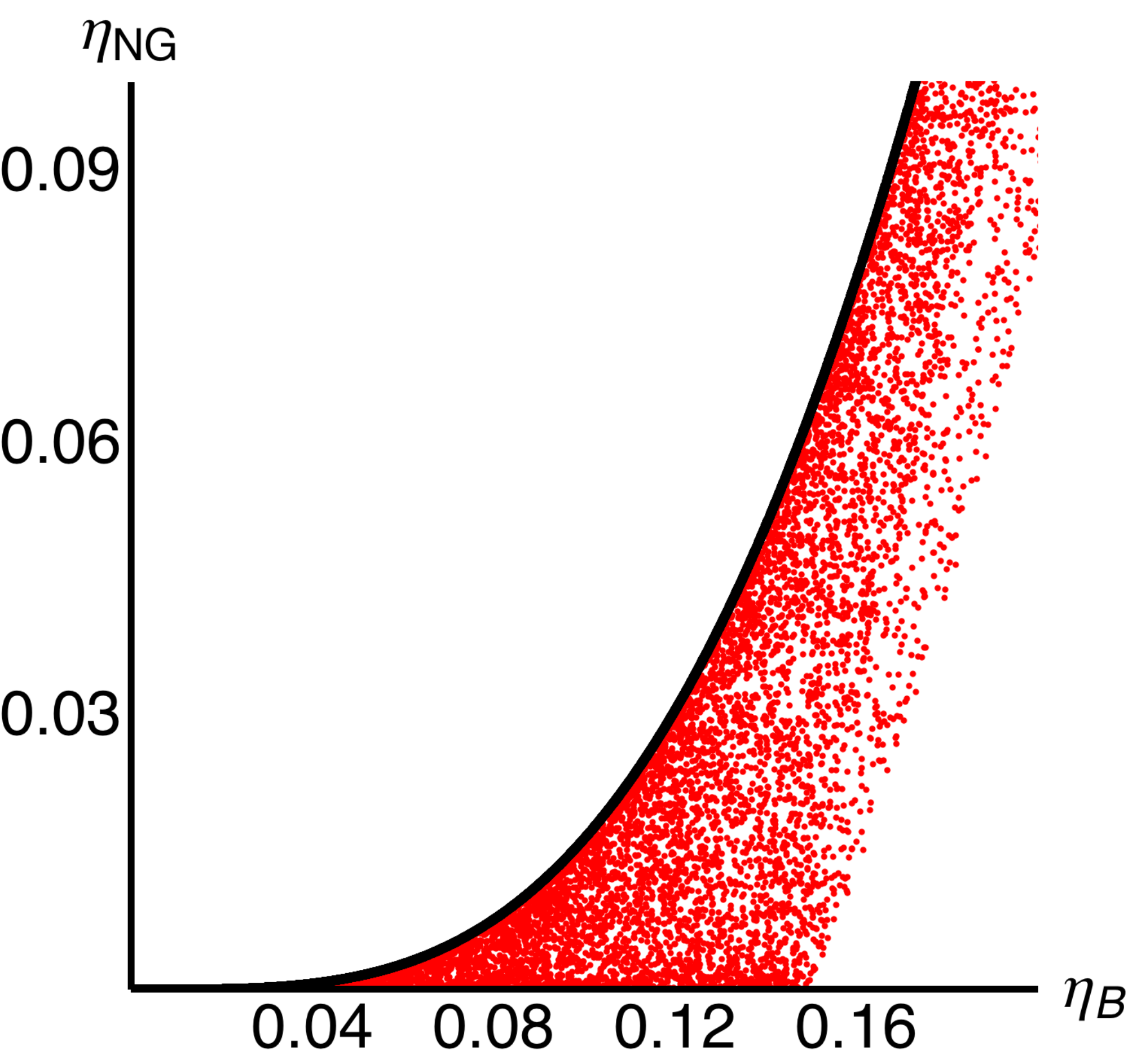}
\caption{Nonlinearity measures for a weakly perturbed harmonic
potential. In the left panel we show  $\eta_\sg [V]$ as
a function of $\eta_\sb [V]$ for random values
of $\epsilon_3 \in [-0.1,0.1]$ and $\epsilon_4 \in [-0.25,0.25]$.
The right panel shows
the corresponding values for
$\epsilon_3 \in [-0.2,0.2]$, $\epsilon_4 \in [-0.25,0.25]$.
The solid black line in both plots is the function 
$\eta_\sg (\eta_\sb)$ of Eq. (\ref{param}).
}
\label{f1}
\end{figure}
\par
%%%%%%
If the nonlinearity is induced by an anharmonic potential that is
an even function at lowest orders, i.e.  corresponds to $\epsilon_3=0$, 
then we have $\alpha_1=0$ and the two measures are monotone of each other, and thus
equivalent in assessing the nonlinearity. More precisely, upon inserting
Eq. (\ref{alphas}) in the expressions of $\eta_\sb$ and $\eta_\sg$, we
have 
\begin{equation}
\eta_\sg = h\left(\frac12 \sqrt{1+24\eta_\sb^2 (\eta_\sb^2-2)}\right)\,.
\label{param}
\end{equation}
On the other hand, if the anharmonic potential contains odd terms, 
then monotonicity is no longer ensured and should be checked for each
specific case. %%%
This behavior is illustrated in Fig. \ref{f1} where
we show parametric plots of $\eta_\sg$ as a function of $\eta_\sb$ 
for potentials corresponding to random values of the parameters
$\epsilon_3$ and $\epsilon_4$. In both panels the solid black curve is the
function $\eta_\sg (\eta_\sb)$ reported in Eq. (\ref{param}).
In the left panel, the red points correspond to random values
of $\epsilon_3 \in [-0.1,0.1]$ and $\epsilon_4 \in [-0.25,0.25]$, 
whereas in the right 
panel we show the corresponding values
of $\epsilon_3 \in [-0.2,0.2]$ and $\epsilon_4 \in [-0.25,0.25]$.
\par
The general perturbative expansion reported above illustrates that a 
potential function may deviate from the harmonic behavior in many different 
ways and "directions" since, loosely speaking, the space of potential functions 
is infinite dimensional (and this is true even restricting attention to the ring of polynomials). As a consequence, one would in principle expect that the nonlinearity 
of an oscillator needs a set of parameters to be characterized. On the other hand, as
we will see in the next Section, we  prove that for a set of relevant potentials our 
measure is indeed capturing and quantifying the intuitive notion of nonlinearity, 
including also cases where the nonlinearity is strong.
%%%%
\section{Examples and discussion}
\label{s:pots}
%3
In this section we evaluate the nonlinearity of some quantum 
oscillators subject to potentials chosen on the basis of their 
relevance, properties, and analytic solvability. We employ the 
results to compare the behavior of the two measures, and to validate
the use of $\eta_\sg$. We consider only position-dependent 
potentials $V(x)$, and confine our investigation to the one-dimensional case,
where, assuming $m=1$ and $\hbar=1$, the Schr\"{o}dinger equation reads
\begin{equation}
\left[-\frac12 \frac{d^2}{dx^2}+V(x)\right]\phi(x)=E\,\phi(x)\,.
\label{eq:sch}
\end{equation} 
In the following, we are going to address in some detail the
nonlinearity of the Morse (M) potential \cite{mor29}, $V_\sm (x)$, the modified 
P\"{o}schl-Teller (MPT) potential \cite{nie78}, 
$V_\smpt (x)$, the modified isotonic oscillator (MIO) potential \cite{car08}, 
$V_\stpt (x)$, and the Fellows-Smith supersymmetric partner
of the harmonic oscillator \cite{fs11} $V_\sfs(x)$.
These potentials describe a wide range of different physical 
systems, with striking different physical properties. 
A common feature, though, is that 
the eigenfunctions depend explicitly on a parameter that couples the 
range of the potential and the depth of the well. For the Morse and 
the MPT potential, this parameter sets also the number of bound states.
%%%%%
\begin{figure}[h!] 
\includegraphics[width=0.44\columnwidth]{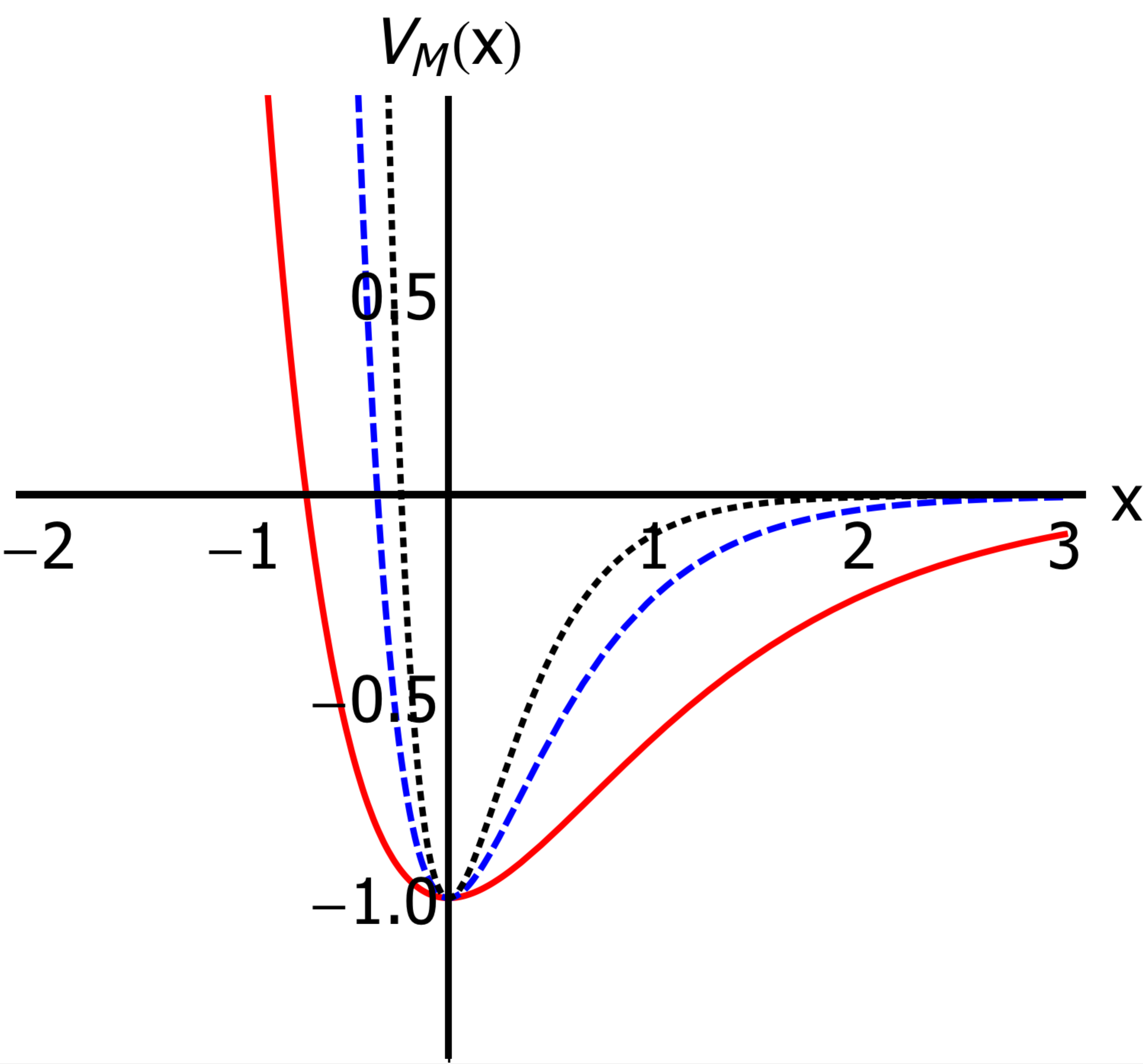}
\includegraphics[width=0.44\columnwidth]{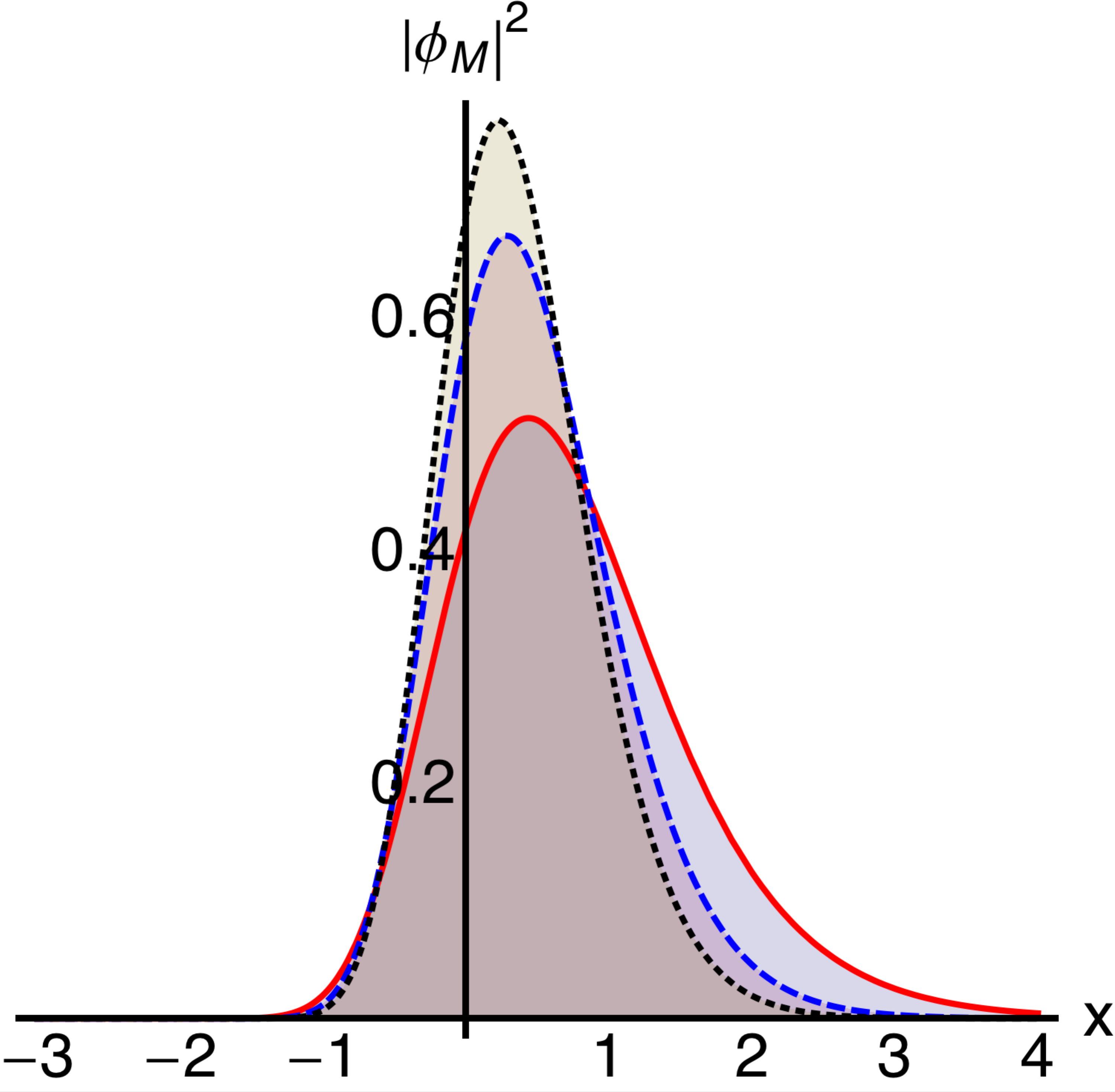}
\includegraphics[width=0.76\columnwidth]{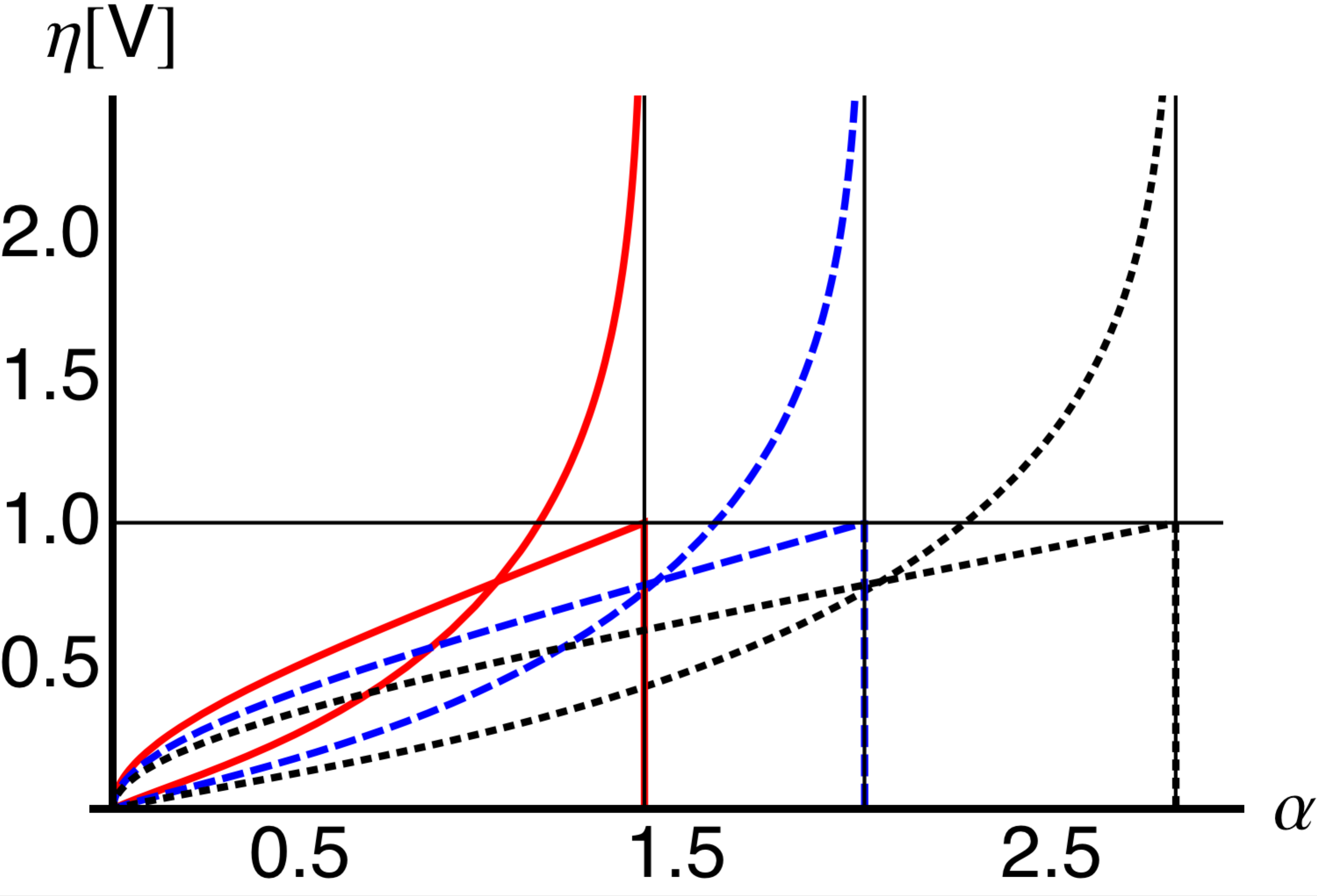}
\caption{The Morse potential. In the upper left panel we show $V_\sm (x)$ for 
$\alpha=1$ (solid red),  $\alpha=2$ (dashed blue), $\alpha=3$ (dotted 
black) and for a 
fixed depth parameter $D=1$.  In the upper right panel we show the 
corresponding GS probability densities, $|\phi_\sm|^{2}(x)$.
In the lower panel we show the nonlinearity measures 
$\eta_\sg\left[\text{V}\right]$ and
$\eta_\sb\left[\text{V}\right]$ as a functions of
$\alpha$ for different values of $D$. We have $D=0.25$ (solid red),
$D=0.5$ (dashed blue), $D=1$ (dotted black). The vertical black lines
are placed in correspondence of the limiting values of $\alpha$.}
\label{f:morse}
\end{figure}
%%%%%%%%%%
\subsection{The Morse potential}
The Morse potential was first suggested by Morse
\cite{mor29} as an anharmonic potential to describe covalent 
molecular bonding. It is an asymmetric potential and its 
expression reads as follows
\begin{equation}
V_\sm(x) = D (e^{- 2 \alpha x}-2 e^{-\alpha x})\,, 
\label{eq:morse}
\end{equation}
where the coordinate $x$ corresponds to the distance from the minimum of the potential. 
The coefficient $D>0$ represents the bond dissociation energy, whereas the 
parameter $\alpha$ controls the width and skewness 
of the well. Indeed, the eigenvalues of the Morse potential
matches very well the experimental spectral lines for the vibration of the
nuclei in diatomic molecules.
The harmonic limit at fixed $D$ is achieved for $\alpha\rightarrow 0$, 
whereas the reference harmonic potential corresponds to a frequency
$\omega_\sr=\sqrt{2D}\alpha$. The form of the potential is
illustrated in Fig. \ref{f:morse} for different values of $\alpha$.
The number of bound states is finite, and is given by the integer part of 
$N= -\frac{1}{2}+\sqrt{2 D}/\alpha$. We thus have the constraint
$\alpha<2 \sqrt{2 D}$ on the parameters in order to have at a least 
one bound state. For vanishing $D$ or for $\alpha$ approaching $2\sqrt{2 D}$ 
there is just one bound state, the GS. 
The GS wavefunction is given by
\begin{align}
\phi_\sm (x) =& \sqrt{2} (2N+1)^N \sqrt{\frac{N \alpha}{2N!}}
e^{-\alpha x N-(N+\tfrac12) e^{-\alpha x}}
\label{eq:eigenMorse}
\end{align}
and corresponds to a bound state with energy
$E_\sm = -\frac12 \alpha N^2$. 
\par
After looking at the shape of the potential in Fig. \ref{f:morse}, one
would expect that the nonlinearity vanishes for $\alpha\rightarrow 0$,
and increases with $\alpha$ at any fixed value of $D$. Indeed, this
intuitive behavior is captured by both measures,
$\eta_\sg\left[\hbox{V}\right]$ and
$\eta_\sb\left[\hbox{V}\right]$, as they grow continuously and
smoothly from zero to the maximum value of $\alpha$ for which the
condition on the existence of bound states is fulfilled. The
nG-based nonlinearity $\eta_\sg[V]$ diverges as
$\eta_\sg[V]\simeq 1+ \frac14 \log (D/4) + \frac12 \log
(\alpha-2\sqrt{2D})$ for $\alpha$ approaching $2\sqrt{2D}$ and vanishes
as $\eta_\sg[V]\simeq y(1-\log y)$, where $y=\alpha/16
\sqrt{2D}$, for vanishing $\alpha$.
At fixed value of $\alpha$ both measures of nonlinearity decrease
with increasing $D$, a behavior that correctly captures
the shape of the potential (which indeed appears {\em more harmonic} when becoming
deeper at fixed width).
%%%%
\subsection{The modified P\"{o}schl-Teller potential}
\begin{figure}[h!]
\includegraphics[width=0.48\columnwidth]{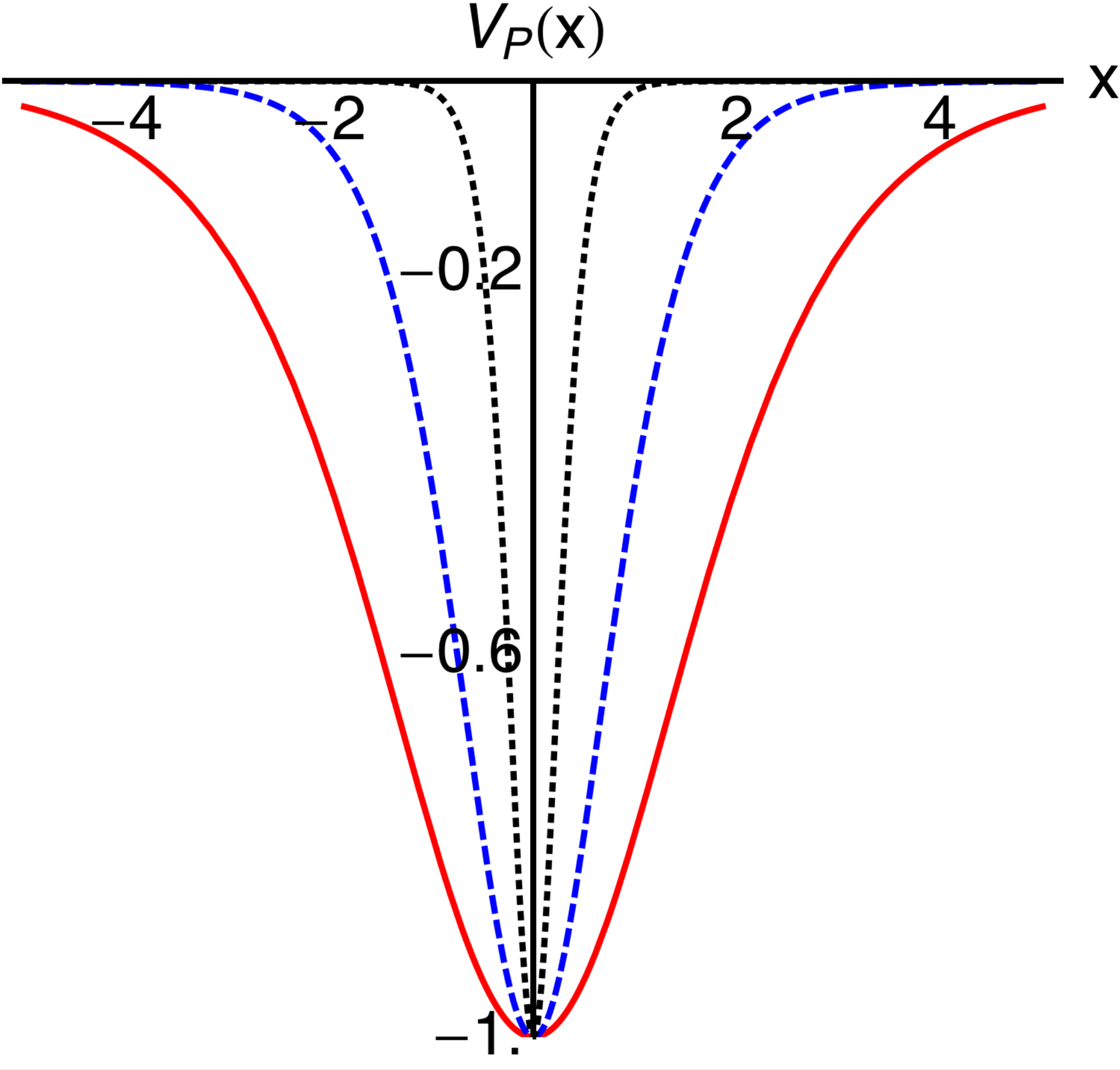}
\includegraphics[width=0.48\columnwidth]{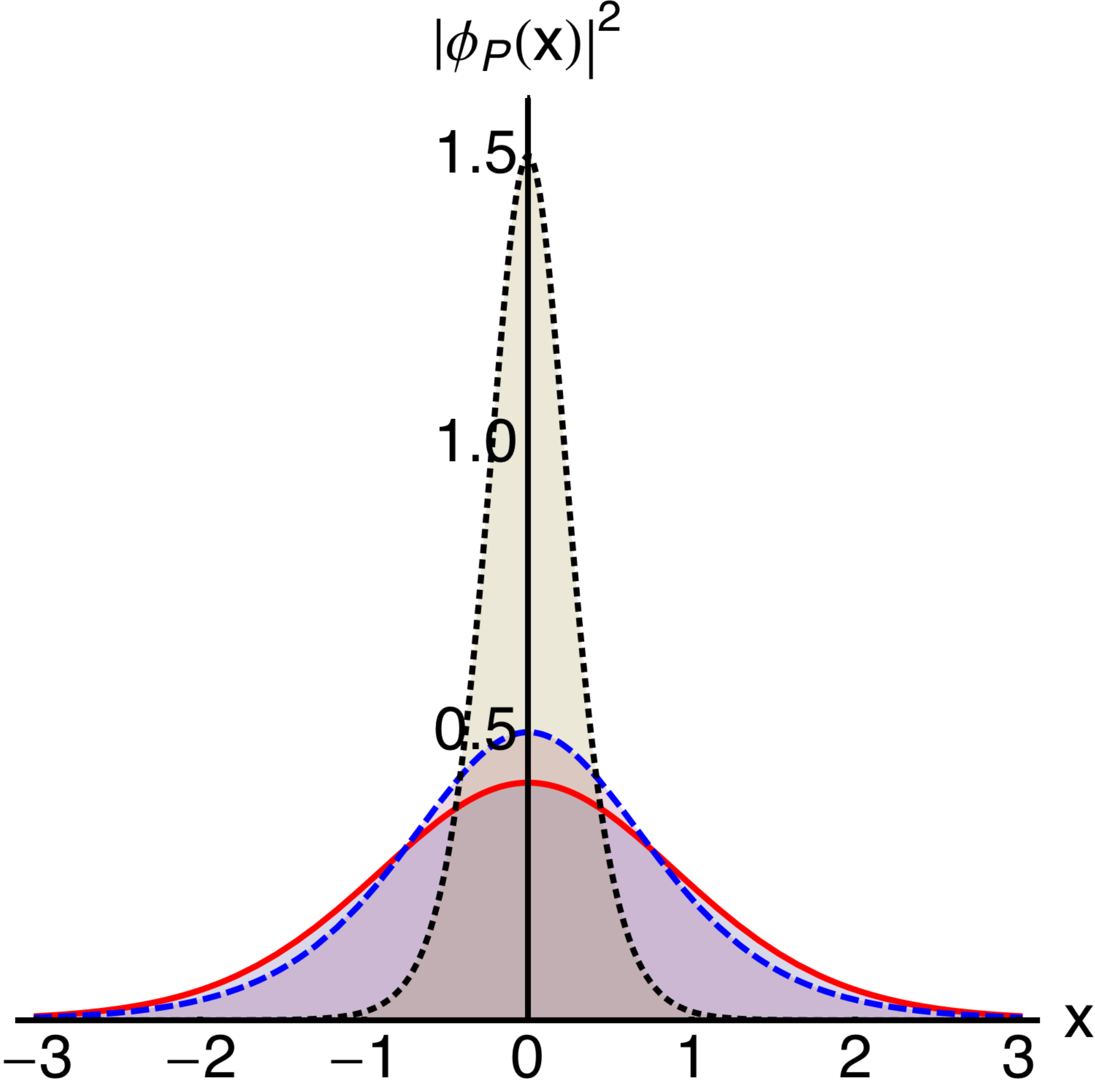}
\includegraphics[width=0.77\columnwidth]{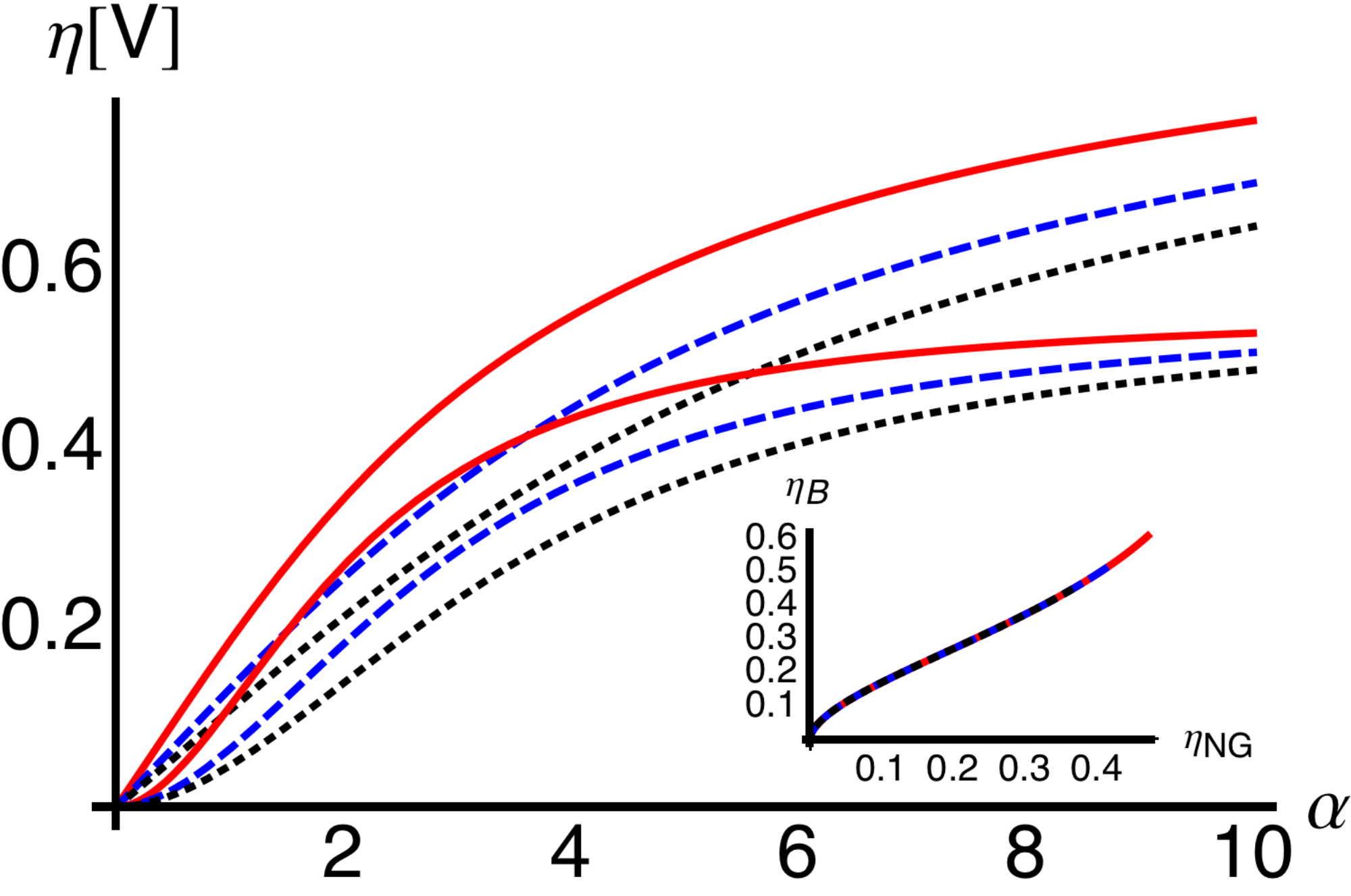}
\caption{The modified P\"{o}schl-Teller potential. The
upper panels show the MPT potential $V_\smpt (x)$ 
and the corresponding
ground state probability density $|\phi_\smpt (x)|^2$ 
for $\alpha=1/2$ (solid red lines), 
$\alpha=1$ (blue dashed) and $\alpha=3$ (black dotted) and a fixed 
potential depth $D=1$. In the lower 
panel we show the nonlinearity measures
$\eta_\sb[V_\smpt]$ and $\eta_\sg[V_\smpt]$ as 
a function of $\alpha$ for $D=1$ (red solid line), $D=2$ (blue dashed),
$D=3$ (black dotted). The inset is a parametric plot of $\eta_\sb$
as a function of $\eta_\sg$, showing that the two measures are monotone
functions of each other, {\em independently} on the value of $D$.}
\label{fmpt}
\end{figure}

The modified P\"{o}schl-Teller potential (MPT) describes several types 
of diatomic molecules bonding. It also appears in the solitary wave 
solutions of the Konteweg-de Vries equation, and finds application 
in the analysis of confined systems as 
quantum dots and quantum wells. 
The modified P\"{o}schl-Teller potential \cite{nie78} is an even 
function, given by 
\begin{equation}
V_\smpt (x)=-\frac{D}{\cosh^2(\alpha x)}\,,
\label{eq:mod}
\end{equation}
where $D>0$ is the potential depth and $\alpha$ is connected to the range 
of the potential. As it will be apparent in the following, it is convenient 
to reparametrize the potential expressing the depth parameter as 
$D=\frac 12 \alpha^2 s (1+s)$, where $$s=\frac12
(-1+\sqrt{1+8D/\alpha^2})>0\,.$$ The harmonic limit for 
any fixed value of $D$  is obtained for $\alpha\rightarrow 0$, whereas 
the reference harmonic potential corresponds to a frequency 
$\omega_\sr=\sqrt{2D}\alpha=\sqrt{s(s+1)}\alpha^2$. The form of the
potential is illustrated in the upper left panel of Fig. \ref{fmpt}, where
$V_\smpt (x)$ is shown for $D=1$ and different values of $\alpha$.
The MPT potential is an even function and thus, according to the
arguments of the previous Section, we expect the two measures to be 
monotone functions of each other, at least for small values of $\alpha$.
The wavefunction of the ground state is given by 
\begin{eqnarray} \phi \smpt(x)= \frac1{\pi^{\frac14}}
\sqrt{\frac{\alpha\,\Gamma[\frac12+s]}{\Gamma[s]}}
\frac1{\cosh^s (\alpha x)}\,,
\end{eqnarray} 
where $\Gamma[x]$ denotes the Gamma function, and correspond to a bound
state with energy $E_\smpt=-\frac12\alpha^{2} s^{2}$. The corresponding
probability density $|\phi_\smpt (x)|^2$ is shown in the upper right panel 
of Fig. \ref{fmpt} for different values of $\alpha$.
\par
In the lower panel of Fig. \ref{fmpt} we show the two nonlinearity measures as
functions of $\alpha$, for $D=1/2$ (red solid line), $D=1$
(blue dashed), $D=3$ (black dotted).  Both measures increase
monotonically with $\alpha$ and decrease with $D$. 
The two measures are anyway monotone of each other 
{\em independently}  on the value of $D$. This is illustrated
in the inset of the lower panel, where we show a parametric plot of $\eta_\sb$
as a function of $\eta_\sg$. The plot has been obtained by varying
$\alpha$ at fixed values of $D$ (the same values used above).
As it is
apparent from the plot, the three curves superimpose each other.
%%%%
%%%%
\subsection{The modified isotonic potential}
\begin{figure}[h!]
\includegraphics[width=0.48\columnwidth]{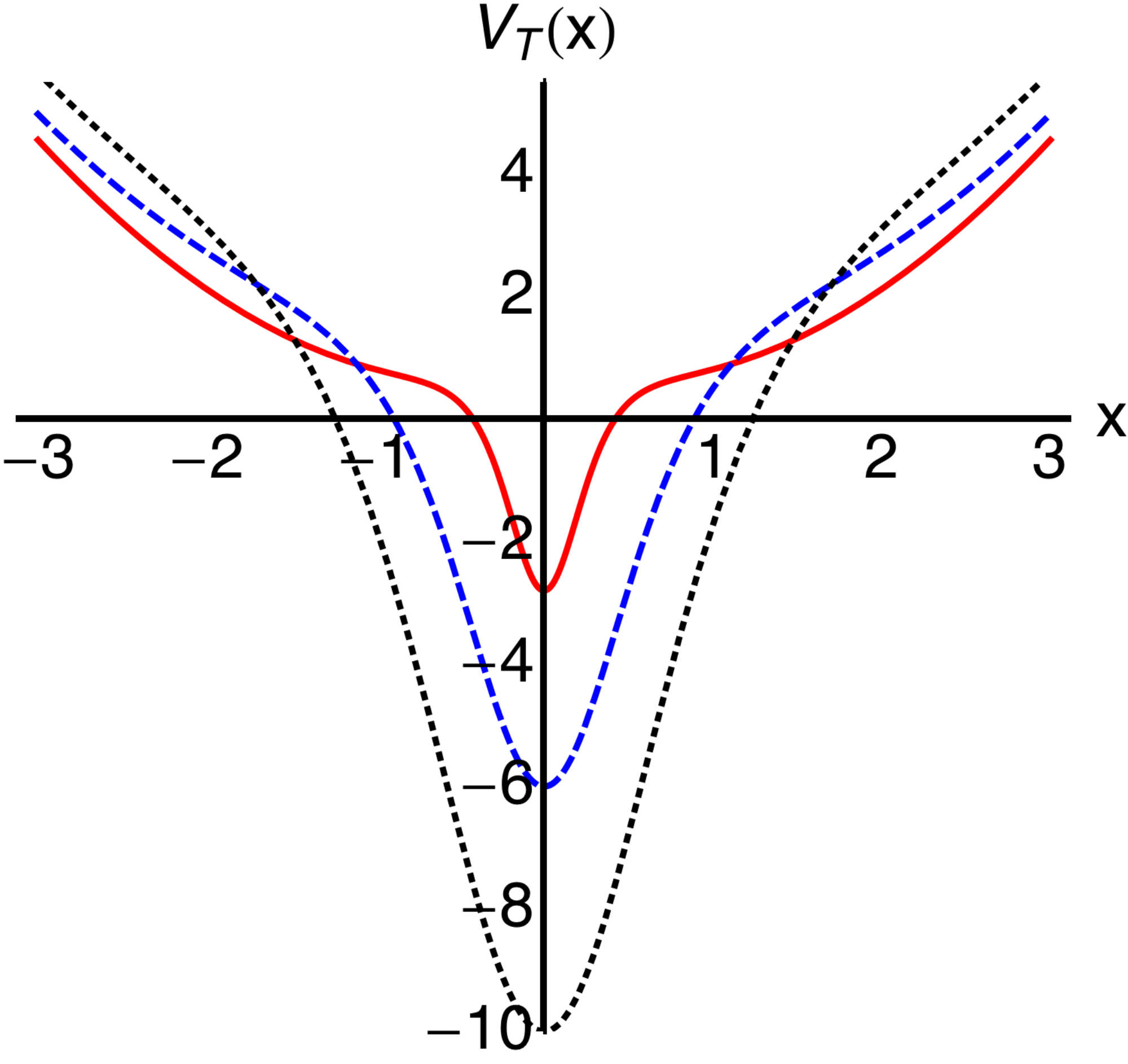}	
\includegraphics[width=0.48\columnwidth]{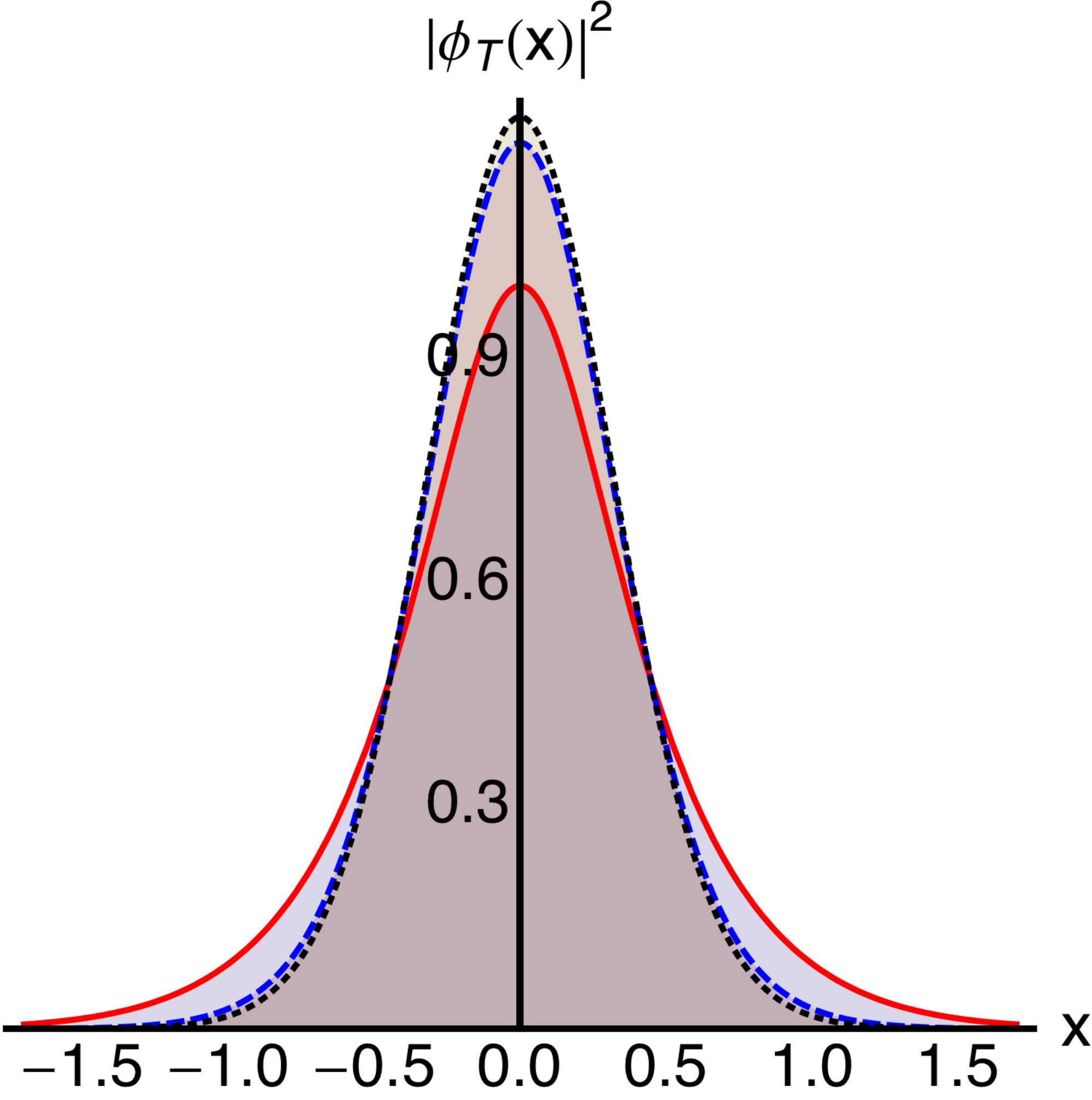}
\includegraphics[width=0.77\columnwidth]{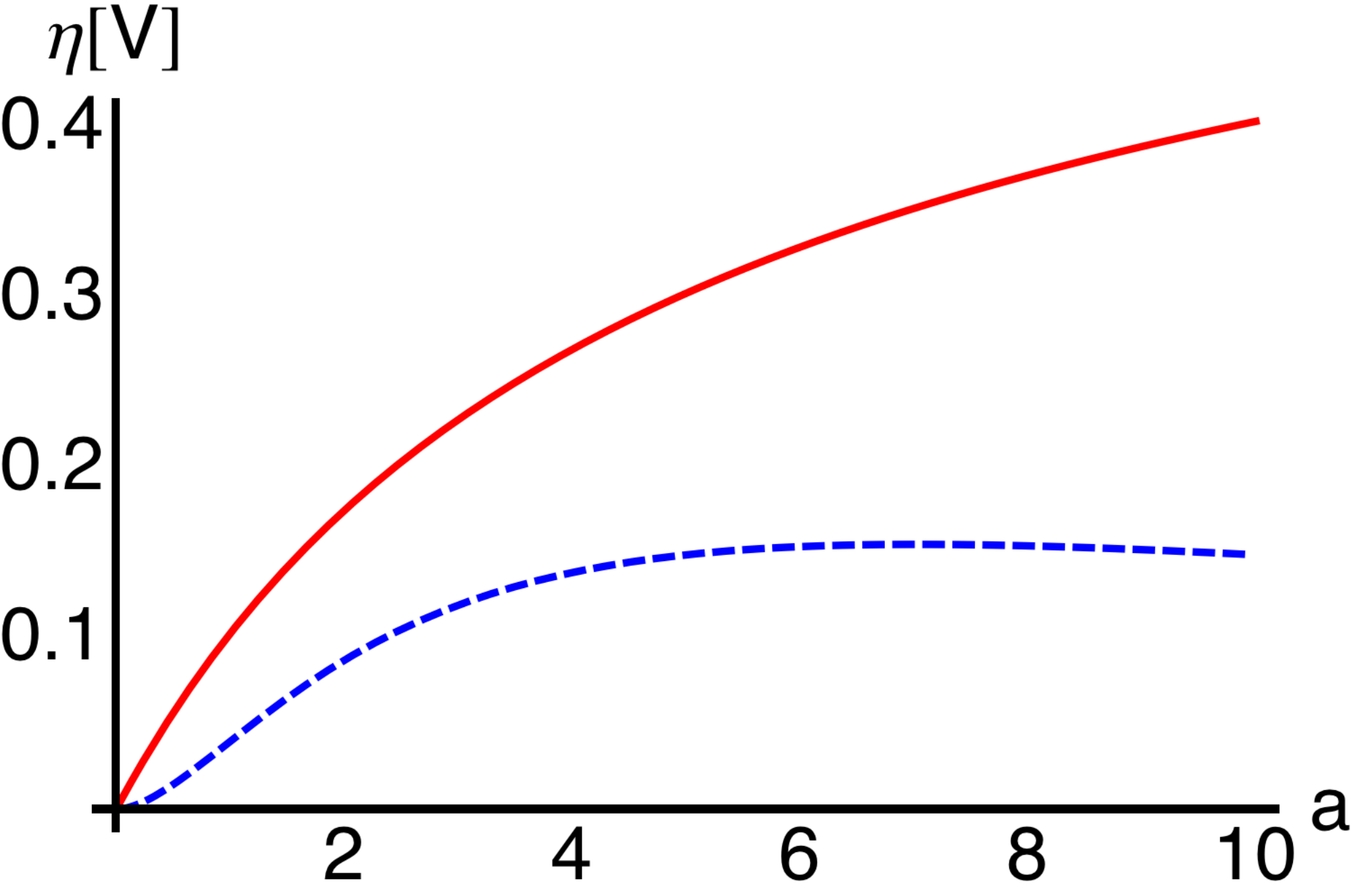}
\caption{The modified isotonic potential. The upper left panel shows the 
MIO potential $V_\stpt(x)$ for different values of the parameter: $a=5$ (solid red), 
$a=1$ (blue dashed) and $a=\frac12$ (black dotted). The upper right panel
shows the probability density $|\phi_\stpt (x)|^2$ 
of the corresponding GS wavefunctions.
In the lower panel we show the nonlinearity measures 
$\eta_\sb \left[\text{V}\right]$ (solid red line) 
and $\eta_\sg \left[\text{V}\right]$ (dashed blue) as a function of 
$a$.} 
\label{mio}
\end{figure}

The so-called isotonic oscillator is a quantum system subjected to 
a potential of the form $V(x) \propto \omega^2 x^2 + g/x^2$ with 
$g>0$. Roughly speaking, the potential is aimed to describe harmonic 
oscillators in the presence of a barrier. The isotonic oscillator has an equally 
spaced  spectrum and it is exactly solvable. 
The potential of the so-called modified isotonic oscillator (MIO)
\cite{car08,yes11} 
is given  by 
\begin{align}
V_\stpt(x) = \frac12 \left[ x^2 + 4 \frac{(a+2) (a x^2-1)}{a (a x^2+1)^2}\right]
\qquad a>0\,.
\end{align}
The MIO class describes a family of oscillators 
which interpolate between 
the harmonic and the isotonic oscillator, and represents a good testbed for a measure
of nonlinearity. We have $V_\stpt(x) \simeq -D + \frac12 \omega_\sr^2 x^2$ for 
small values of $x$, where the depth of the potential is given by
$D=2(a+2)/a$ and the reference frequency by $\omega_\sr=\sqrt{25 +12a}$,
whereas for large $x$ the potential approaches $V_\stpt(x) \simeq \frac12 x^2
+ D/(a x^2)$. The form of the potential for different values of $a$ is
shown in the left panel of Fig. \ref{mio}.
The ground state has energy $E_\stpt=\frac12 - \frac4a$ and the
corresponding wavefunction is given by 
$$
\phi_\stpt (x) = \frac1{\pi^\frac14
\sqrt{\Phi(\frac4a,\frac12+\frac4a;\frac1a)}}\, 
e^{-\frac12 x^2} \left(\frac1a + x^2\right)^{-\frac{2}{a}}\,,
$$
where $\Phi(a,b;z)$ is the confluent hypergeometric function.
As the form of the potential may suggest, oscillators subjected to 
MIO potentials have the ground state detached from the rest of 
the eigenstates, which are equally spaced in energy.
The  GS probability densities, $|\phi_\stpt|^{2}(x)$ is shown 
in the upper right panel of Fig. \ref{mio}.
\par
We have evaluated  the nonlinearity measures as a function of the 
parameter $a$ and the results are reported in the lower panel
of Fig. \ref{mio}. As it is apparent from the plot, the two measures 
are monotone of each other, and may be used equivalently, as far as the value
of $a$ is not too large. For increasing $a$ the Bures measure continue
to grow whereas $\eta_\sg$ has a maximum and then starts to decrease,
thus no longer representing a suitable quantity to assess 
the nonlinear features of $V_\stpt$. This behavior is due to the
peculiar structure of the eigenstates: Indeed the ground state 
of the system, though departing from that of the harmonic reference,
is becoming more and more Gaussian, thus resembling that of a harmonic
oscillator (not the reference one). 
In fact, the nonlinear features of the potential are encoded
in the rest of the eigenstates. In order to capture the nonlinear
features of this kind of potentials, we have to use $\eta_\sb$ or
to look at the non-Gaussian properties of states at thermal equilibrium,
which account for the whole spectrum. 
%%%%%%%%%%%%%
%%%%%%%%%%%%%
\subsection{The Fellows-Smith potential}
\begin{figure}[h!]
\includegraphics[width=0.48\columnwidth]{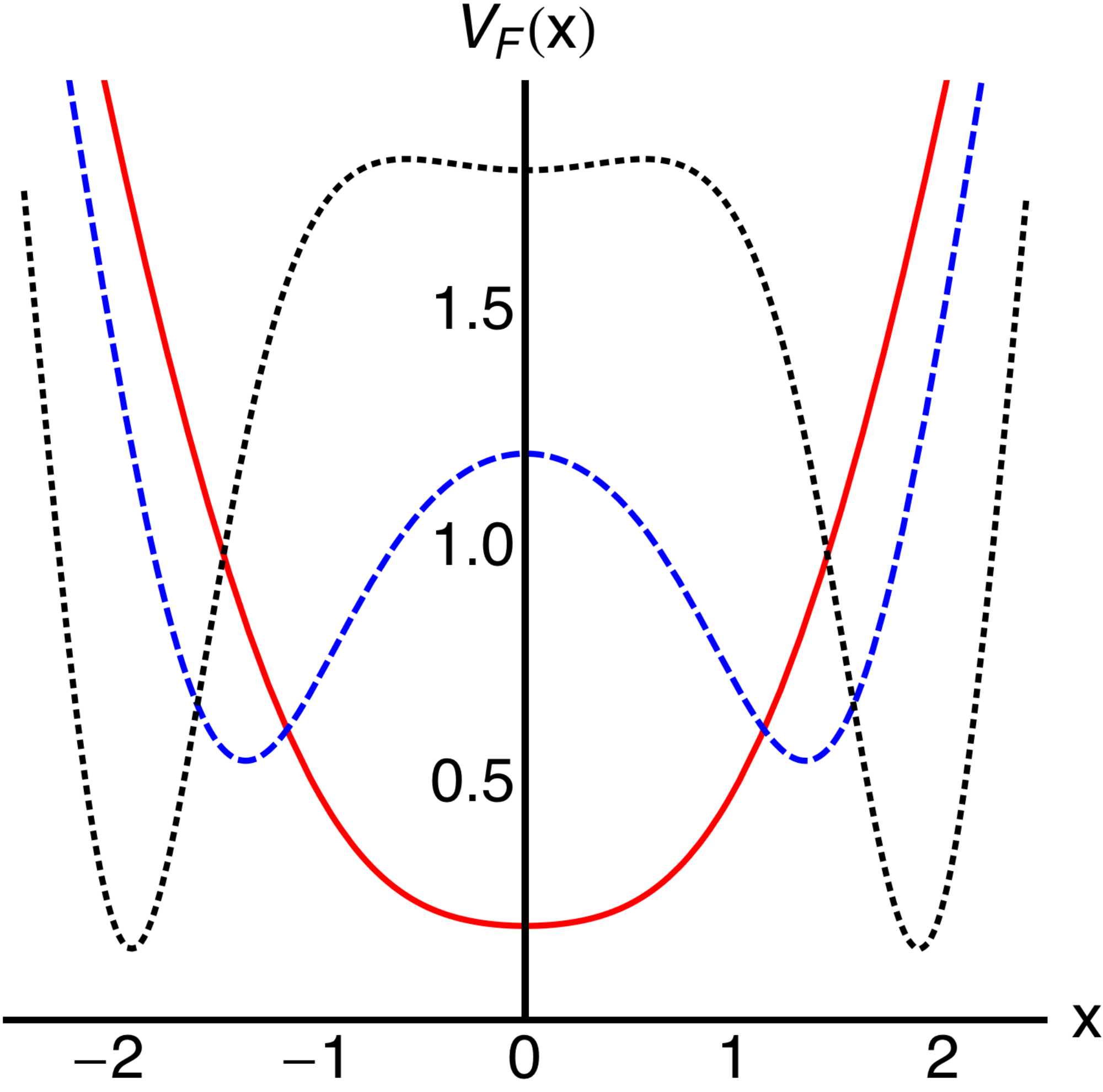}	
\includegraphics[width=0.48\columnwidth]{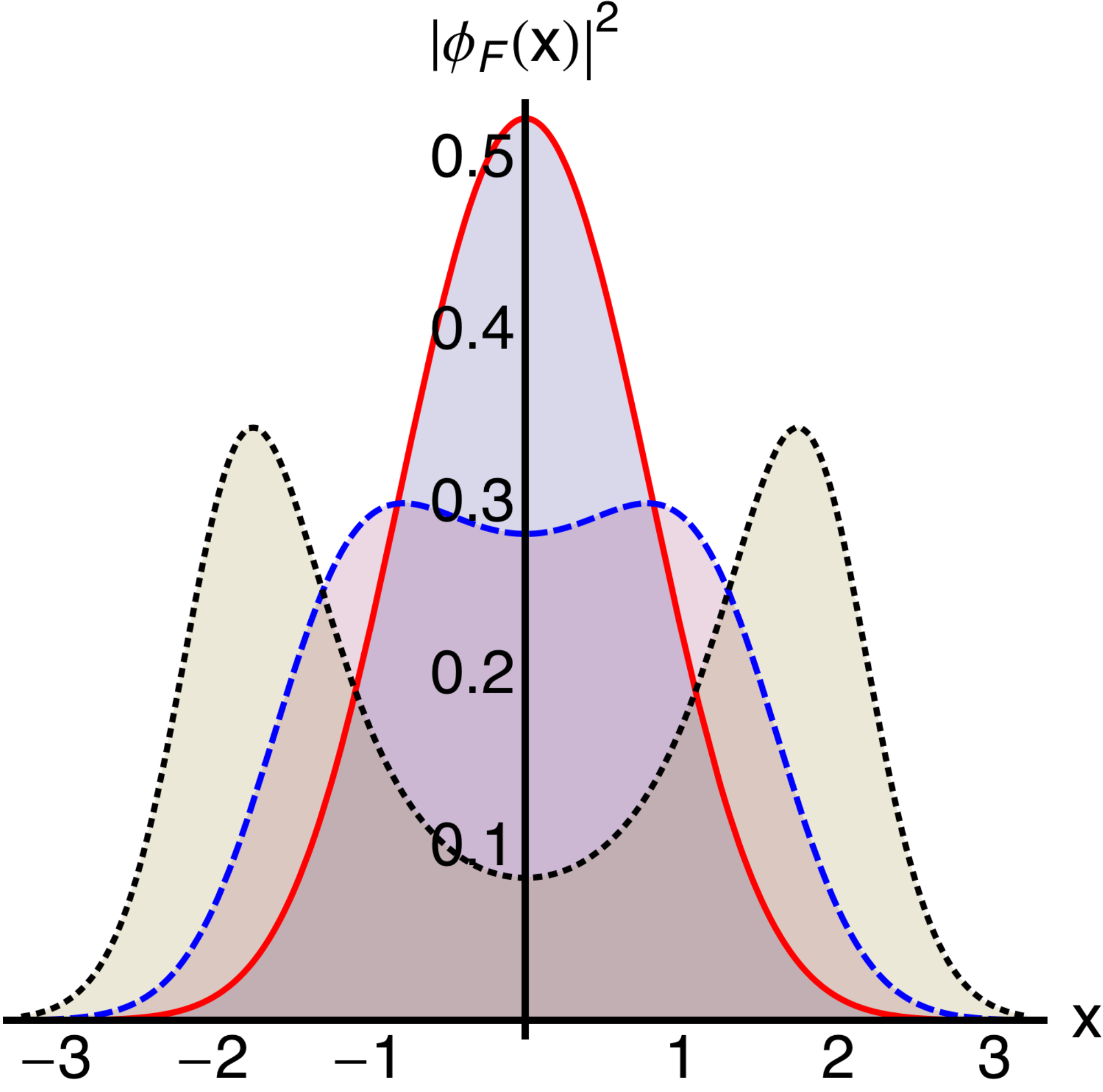}
\includegraphics[width=0.72\columnwidth]{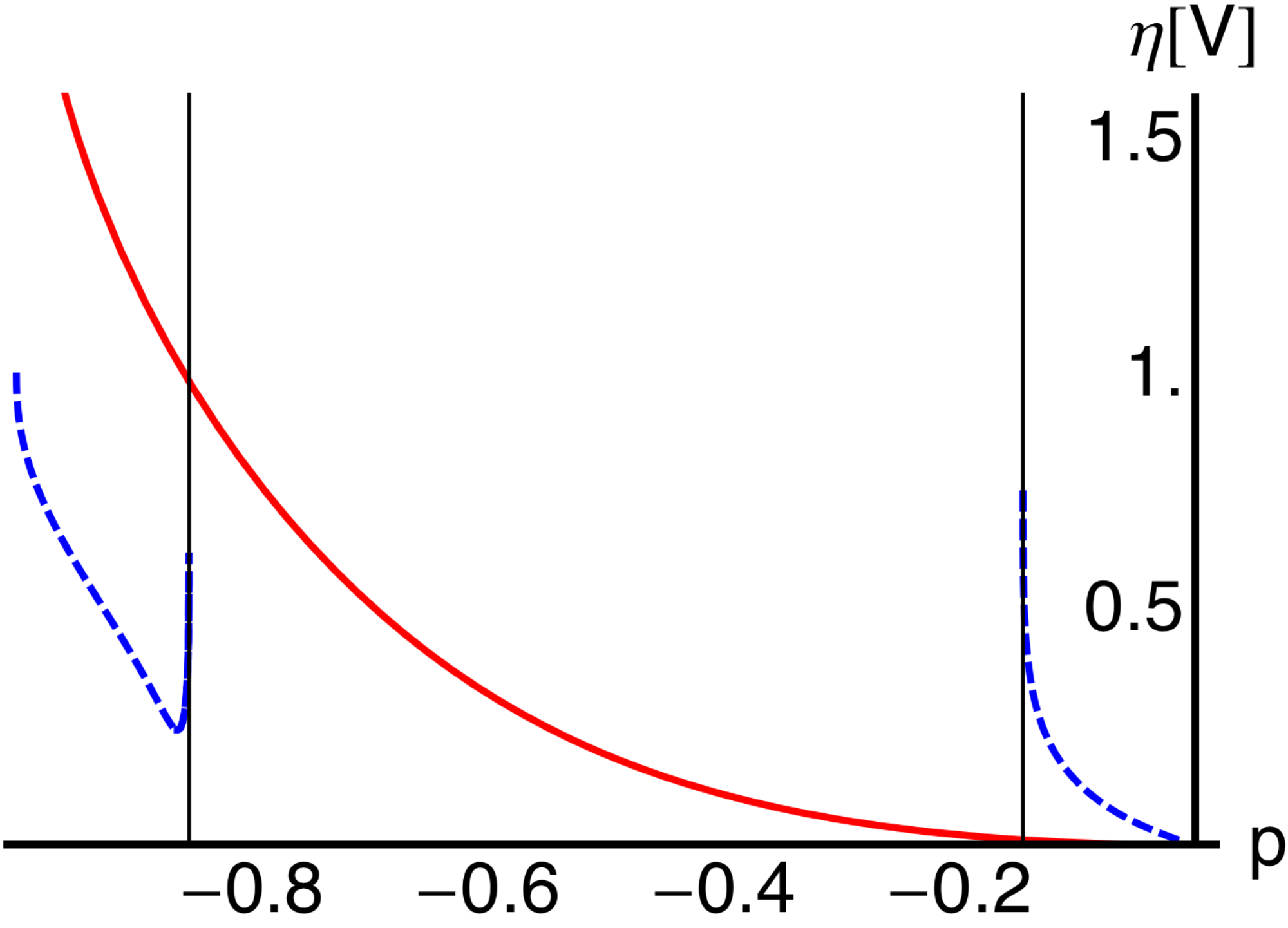}
\caption{The Fellows-Smith potential. The upper left panel shows
$V_\sfs(x)$ for different values of the parameter: $p=-\frac1{10}$ 
(solid red), $p=-\frac25$ (blue dashed) and $p=-\frac{9}{10}$ 
(black dotted). The upper right panel
shows the probability density $|\phi_\sfs (x)|^2$ 
of the corresponding GS wavefunctions.
In the lower panel we show the nonlinearity measure 
$\eta_\sg \left[\text{V}\right]$ (solid red line) as a function of
$p$, together with the measure
$\eta_\sb \left[\text{V}\right]$ (dashed blue) in the regions where it
can be evaluated. The vertical black lines denote the values $p=p_\pm$. 
} 
\label{figfs}
\end{figure}

We end this Section by considering a set of nonlinear oscillators
corresponding to a
class of potentials which have no clear harmonic reference, such that 
the Bures measure of nonlinearity cannot be properly defined. 
These correspond to a class
of supersimmetric partners of the harmonic potential, given by
\cite{fs11}
\begin{align}
V_\sfs (x) &= -2p + \frac12 x^2  + 4\, (1+p)\, x^2
\frac{\Phi(\frac{3+p}{2},\frac32;x^2)}{\Phi(\frac{1+p}{2},\frac12;x^2)^2}
\notag \\ \times &
\left[ (1+p)\,
\Phi(\frac{3+p}{2},\frac32;x^2)-\Phi(\frac{1+p}{2},\frac12;x^2)
\right]\,, 
\end{align}
where $\Phi(a,b;z)$ is the confluent hypergeometric function and
$p\in(-1,0]$. The potentials show a single well structure for 
$p\in [p_+,0]$, a double well
structure for 
$p\in [p_-,p_+]$ and a triple one for
$p\in [-1,p_-]$, where $p_\pm=-\frac12\pm\frac{\sqrt{2}}{4}$.
The behavior of $V_\sfs(x)$ in the different regions is illustrated in 
the upper left panel of Fig. \ref{figfs}, where we show the potentials for 
$p=-\frac{1}{10},-\frac35,-\frac9{10}$ respectively. The corresponding 
probability distributions of the ground state $|\phi_\sfs (x)|^2$
are shown in the upper right panel of the same figure. The wave function of
the ground state is given by
$$
\phi_\sfs(x) = \frac{1}{\pi^{\frac14}}\,\sqrt{\frac{2^p}{\Gamma[1+p]}}\,
\frac{\Gamma[1+\frac{p}2]\, e^{\frac12
x^2}}{\Phi(\frac{1+p}{2},\frac12;x^2)}\,,
$$
and corresponds to the eigenvalue $E_\sfs=\frac12-p$.
\par
As it is apparent
from the plot and from the structure of the potential, it is possible to
define a proper reference harmonic oscillator only 
for $p\in [p_+,0]$: in this case we have $\omega_\sr =
\sqrt{1+8p(1+p)}$, which is vanishing for $p\rightarrow p_\pm$. 
For $p\in [p_-,p_+]$ there is no such option, unless
one breaks the symmetry of the potential and arbitrarily  choose one of the
two minima to define a reference harmonic oscillator. In the third
region, $p\in (-1,p_-]$, the potential shows again a  mininum at $x=0$. 
However, using this feature to define the reference harmonic potential 
is obviously misleading, since it ignores the main features of the potential.
%%%%
\par
%%%%%%%%%%%%%
We have evaluated the nonlinearity measure $\eta_\sg$ for $p\in(-1,0]$
and the Bures one $\eta_\sb$ where it is possible. Results are shown
in the lower panel of Fig. \ref{figfs}, where the solid red line 
and the blue dashed one denote respectively $\eta_\sg$ and $\eta_\sb$. 
As it is apparent from
the plot, $\eta_\sg$ monotonically decreases with $p$, thus properly 
capturing the nonlinear behavior of the $V_\sfs$ potential. Besides, 
the two measures are monotone of each other for $p\in [p_+,0]$, where the 
Bures measures may be properly defined. The plot also makes apparent
that despite $\eta_\sb$ may be calculated also for $p\in (-1,p_-]$, its
behavior is not consistent with the behavior at smaller values of 
$p$, thus failing to provide a quantitative assessment of nonlinearity.
In particular, $\delta_\sb\rightarrow 1$ for $p\rightarrow p_-$,
then, as $p$ decreases, it shows a minimum and then starts to 
increase for $p$ that decreases to $p=-1$.
%%%%%%%%%%%%%
\section{Conclusions}
\label{s:out}
Quantum oscillators with nonlinear behavior induced by anharmonic 
potentials have attracted interest in different fields, as they play 
a relevant role for fundamental and practical purposes. In
particular, they have recently received attention as a possible
resource for quantum technology and information processing. 
As a consequence, it would be useful to have a suitable measures of nonlinearity 
to better characterize these potentials and to better assess their performances 
in those fields. 
In this paper, we have addressed the quantification 
of nonlinearity for quantum oscillators, and have introduced
two measures of nonlinearity based on the properties of the ground state of 
the potential, rather than on the form of the potential itself. 
The first measure accounts for the Bures distance between the potential
GS and that of a reference harmonic potential. It is a natural choice
for a measure of nonlinearity, however it requires the knowledge of the
potential near its minimum. We have thus suggested a different measure, based
on the non-Gaussian properties of the potential GS, which may be
calculated using only information about the GS itself.
\par
The two measures 
have been analyzed and compared, both in terms of their general
properties and by evaluating them for some significant anharmonic 
potentials. Our results show that the nG-based measure has some 
merits which makes it a good choice for the purpose of assessing
nonlinearity. In fact, while it captures the nonlinear features
of oscillators as the Bures measures in most cases  (e.g., for any
anharmonic potential which is an even function at lowest orders)
it has a clear advantage from a 
computational and experimental point of view: it does not require
the determination of a reference frequency and thus it does not need 
any {\em a priori} information on the corresponding potential to be
calculated. The only ingredient needed to evaluate $\eta_\sg$ is the GS
wavefunction, a task that can be pursued by tomographic 
reconstruction independently on the specific features of the potential.
In addition, we have seen examples of potentials where the Bures measure
cannot be defined, due to the lack of a proper reference harmonic
potential, whereas the nG-based properly quantify the nonlinear
features of the oscillator behavior.  {Moreover, from a more
fundamental point of view, the validity of the measure $\eta_\sg$ is 
strengthened by the fact that the same amount of non-linearity is 
assigned to non-linear Hamiltonians which are related by
symplectic transformation in phase-space (displacement, phase-rotation
and squeezing), inducing a reasonable and expected hierarchy.}
\par
Overall, we have addressed the general issue of assessing the 
nonlinearity of a quantum potential, highlighted the current 
limits, and nevertheless individuated a consistent
method to quantify the nonlinearity based on non-Gaussianity of the
potential's ground state. {In order to fully validate the measure(s) here proposed we would have needed an already established way to compare nonlinear potentials and assess their diversity. Then we could have tried to prove some form of continuity of our measure(s) with respect this quantity .Not having a measure or a set of criteria of this kind is among the motivations of our work while being able to summarize the nonlinear character by a single quantity is the main result.}
\par
Finally, we notice that our results could be exploited in any experiments, e.g., on quantum control, where either by  technological of fundamental issues,
information on the confining potential in inaccessible or limited. Our approach may be generalized and refined by taking into account  the non-Gaussian features of the Gibbs thermal states of the nonlinear  oscillators, rather than the sole GS.   
%%%%%%%%
\section*{Acknowledgments}
This work has been supported by the MIUR project 
FIRB-LiCHIS-RBFR10YQ3H.  
NS acknowledges support from UK EPSRC.
BT is supported by the \quotes{ICTP TRIL Programme 
for Training and Research in Italian Laboratories}. MGG 
acknowledges support from UK EPSRC (EP/K026267/1). 
%%%%%%%%
\appendix
%%%%%%
\section{Gaussian states}
\label{apb}
The density operator of a single-mode continuous-variable state $\rho$ can 
be fully represented by its characteristic function,
\begin{equation}
\chi[\rho](\mathbf{\lambda})=\Tr[\rho D(\lambda)],
\end{equation}
where $\lambda$ is a complex number and $D(\lambda)$ is the
displacement operator $D(\lambda)= e^{\lambda a^\dag - \lambda^* a}$.
Equivalently, we may describe the quantum state using its Wigner function, 
which is the Fourier transform of the characteristic function
\begin{equation}
W[\rho](z)=\int\frac{\hbox{d} 
\lambda^2}{\pi^2}\, e^{\lambda^* z - \lambda z^*}\,
\chi[\rho](\lambda)\,.
\end{equation}
A quantum state is said to be Gaussian if its 
characteristic function (and thus also the Wigner function) is 
Gaussian. Before writing the expression explicitly, we must introduce the 
vector of mean values $\bar X$ and the covariance matrix 
$\sigma$, with elements 
\begin{align}
\bar X_k &= \langle R_k \rangle \notag \\
\sigma_{jk} &= \frac{1}{2} 
\langle \{R_j,R_k\}\rangle - \langle R_j\rangle \langle R_k\rangle \,,
\end{align}
where $R=(x,p)$, $\{A,B\}=AB+BA$, and $\langle A\rangle =
\Tr\left[\rho A\right]$.
The Wigner function of a Gaussian state $\rho_G$ is equal to
\begin{equation}
W[\rho_G](X)=\frac{1}{2\pi\sqrt{
\hbox{det}[\sigma]}}\hbox{exp}
\left[-\frac{1}{2}(X-\bar {X})^T
\sigma^{-1}(X-\bar{X})\right]
\end{equation}
where $X=(\hbox{Re}\, z,\hbox{Im}\, z)$. 
Gibbs thermal states and ground states of Hamiltonians that are at most bilinear 
in the mode operators are Gaussian states, the harmonic oscillator being a 
paradigmatic example.

\end{document}